\documentstyle[preprint,prb,epsf,aps]{revtex}
\newcommand{\bmp}{{\mbox{\boldmath $p$}}}
\newcommand{\bmr}{{\mbox{\boldmath $r$}}}
\newcommand{\bmq}{{\mbox{\boldmath $q$}}}
\newcommand{\qq}{{|\mbox{\boldmath $q$}|}}

\begin{document}
\preprint {WIS-01/05 April-DPP}
\draft
\date{\today}
\title{Relativistic approaches to structure functions of nuclei}
\author{S.A. Gurvitz$^{1,2}$ and A.S. Rinat$^1$}
\address{$^1$Weizmann Institute of Science, Department of Particle Physics,
Rehovot 76100, Israel\\
$^2$Cyclotron Institute, Texas A$\&$M University,
College Station, Texas 77843, USA}
\maketitle
\begin{abstract}
We employ a propagator technique to derive a new relativistic $1/\qq$
expansion of the structure function of a nucleus, 
composed  of point-nucleons. We exploit non-relativistic features of 
low-momentum nucleons in the target  and only treat relativistically
the nucleon after absorption  of a high-momentum virtual photon. The new 
series permits a 3-dimensional reduction of  each term and a formal
summation of all Final State Interaction terms. We then show that a 
relativistic structure function
can be obtained from its non-relativistic analog by a 
mere change of a scaling variable and  an addition of an energy shift.
We compare the obtained result with an ad hoc generalized 
Gersch-Rodriguez-Smith theory,
previously used in computations of nuclear structure functions.

\end{abstract}
 
\section{Introduction}
 
The major tool for computing nuclear structure functions, as 
measured in inclusive electron scattering on nuclei, is the 
Impulse (or Born) Series (IS) in the residual 
interaction between the struck nucleon and the remaining spectator 
nucleus. The lowest order term of that series is 
the widely used Impulse Approximation (IA).
Higher order, Final State Interaction (FSI)
terms are essential for an accurate calculation
of the data, but their determination in practice constitutes a
formidable problem (see for instance Refs.\cite{cio,om1,os,cio1,koh,braun}).  

In the non-relativistic regime there exists an alternative 
approach, originally proposed by Gersch, Rodriguez and Smith
(GRS)\cite{grs}. 
There the structure  function  is expressed in terms of
commutators involving the residual interaction and 
appears, for fixed values of a scaling  variable 
$y$\cite{grs,west}, as a series in inverse powers
of the 3-momentum transfer $\qq$. That theory has extensively been used
to compute  structure functions (or responses) of quantum
gases\cite{retal}.
 
If convergent, the GRS and the IS, taken
to all orders, obviously produce identical results, but
this is not the case if  these series are truncated at some finite order. 
An issue is then which of the truncated series is
a better approximation to the total structure function. 
Judged by the lowest order terms applied to  classes of exactly solvable models, 
the GRS expansion is  to be preferred over the IS
\cite{koh,pg,gr1,sag1,sag2}.  
 
The availability of data obtained with high energy beams requires a theory
valid for the relativistic regime.  In the IS, Final
State Interactions (FSI) 
are summed by means of a 4-dimensional  scattering operator, 
which satisfies a coupled-channel
Bethe-Salpeter equation, but their solution remains  a complicated,
relativistic many-body problem.
 
As regards GRS, 
no satisfactory relativistic extension of the non-relativistic GRS 
theory has been formulated before. A start has been made by 
one of the authors, who previously  exploited  a propagator technique 
for the desciption of the structure function of
composite systems similar to the one used for non-relativistic 
systems.  Formally exact expressions have been derived 
for  relativistic structure functions \cite{sag1,sag2} in terms of
4-dimensional integrals over relativistic propagators and
scattering operators, as is the case in the  
relativistic IS treatment of structure functions. 

In this paper we  develop a 
relativistic GRS series for structure functions exploiting  
manifestly non-relativistic features of the system. There we
shall emphasize that 
only the nucleon which absorbs the virtual photon in inclusive scattering 
acquires a large momentum and has to be treated relativistically. All others  
nucleons  have non-relativistic momenta and can be treated 
accordingly. 

We shall show below that the above non-relativistic features permit an
accurate 3-dimensional reduction of all the terms in the relativistic GRS 
series for a structure function. This feature gives 
the GRS a definite advantage over 
IS. For it  a 3-dimensional  reduction is very involved 
due to negative energy poles in relativistic nucleon propagators.

The outline of this paper is as follows.  In Section II we 
re-derive the non-relativistic GRS series, showing the 
way  to a  relativistic extension,  which is performed 
in Section III. In Section IV we exploit 
non-relativistic features of the problem and 
subsequently prove a 3-dimensional reduction of the lowest order and  
of all higher order FSI terms of the relativistic GRS series.  We 
demonstrate  that the latter can be summed in a closed expression, involving
a 3-dimensional Lippmann-Schwinger, many-channel $T$-matrix. This reduces an 
evaluation of the relativistic nuclear structure functions to a
non-relativistic problem. In the end we relate our final expressions with 
approximative representations of the relativistic GRS series, which have
been used in a description of nuclear structure functions.

\section{Nonrelativistic treatments of structure function}

\subsection{The Impulse Series.}
We start with the non-relativistic structure function  per nucleon
$W(\nu ,\bmq )$, appropriate to a nucleus of $A$ point-nucleons
where $\nu$ and $\bmq$ are the energy and momentum transfered to the target.
In order to simplify the algebra, we restrict our derivation to the case 
of spinless particles. We  focus on the
incoherent part of $W$ which dominates for large $\qq$ and exploit its relation
to the  imaginary part of the forward Compton amplitude
\begin{eqnarray}
W(\nu ,\bmq )&=&-\frac{1}{\pi}{\rm Im}
\langle \Phi_A|{\cal Q}_1^\dagger (\nu, \bmq)G_A(E_A,0)
{\cal Q}_1(\nu, \bmq)|\Phi_A\rangle
\nonumber\\
&\equiv&-\frac{1}{\pi}{\rm Im}
\langle \Phi_A|G_{1,A}(E_A+\nu,\bmq)|\Phi_A\rangle\ . 
\label{a1}
\end{eqnarray}
Here $\Phi_A$ is the ground state wave function of the
target with energy $E_A$, and 
$G_A(E_A,0) =(E_A-H_A)^{-1}$, the exact Green's function 
of the $A$-nucleon system  at rest. 

The operator ${\cal Q}_1({\cal Q}_1^\dagger )$  
shifts the energy and the momentum of a selected  nucleon 
$'$1$'$ by 
$\nu$ and $\bmq$ due to the absorption (emission) of a virtual photon.
The second line in Eq. (\ref{a1}) defines the corresponding
shifted Green's function. The latter is conveniently described, 
using a decomposition of the target
Hamiltonian $H_A$ into a sum of the Hamiltonian $H_{A-1}$
of the $A$-1 nucleon spectator, the 
kinetic energy $K_1$ of a nucleon ($'1'$) and  the 
residual interaction $V_1=\sum_{i\ge 2}V_{1i}$, thus
\begin{eqnarray}
G_{1,A}(E_A+\nu,\bmq )={1\over E_A+\nu-H_{A-1}-K_1(\bmq )-V_1 +i\eta}\ ,
\label{a2}
\end{eqnarray}
where $K_1(\bmq )=(\hat\bmp +\bmq )^2/2M$ is the kinetic
energy operator with the momentum operator  $\hat\bmp$  shifted by $\bmq$
and $M$ is the nucleon  mass. We assume that $NN$ potentials are local, 
$V_{ij}\equiv V_{ij}(\bmr_i- \bmr_j)$. 
Consequently ${\cal Q}_1^{\dagger}(\nu,\bmq)V_1{\cal Q}_1 (\nu, \bmq )=V_1$ 
such that the interaction is not affected by the shift 
as is explicit in Eq. (\ref{a2})). 

At this point we comment on notation. We distinguish
between external parameters as are $E_A, \bmq$ and $\nu$, and
variables which depend on the chosen representation of operators. We 
do not display those variables, unless  required for clarity. 

The most common treatment of the structure function is the Impulse
Approximation (IA), obtained by taking  $V_1\to 0$ in  Eq. (\ref{a2}). 
The shifted Green's function $G_{1,A}(E_A+\nu,\bmq )$ in this 
approximation $G_{1,A}\simeq G_{1,A}^{(0)}$ reads
\begin{eqnarray}
G_{1,A}^{(0)}(E_A+\nu,\bmq ) =
{1\over E_A+\nu-H_{A-1}-(\hat\bmp +\bmq )^2/2M +i\eta}\ . 
\label{a3}
\end{eqnarray}

With a relativistic extension  in mind, we express the above 
$G_{1,A}^{(0)}$ as a convolution of Green's functions for the $(A-1)$-nucleon
spectator and for the struck nucleon ($N$) 
\begin{eqnarray}
G_{1,A}^{(0)}(E_A+\nu ,\bmq )=
i\int {dp_0\over 2\pi}G_{A-1}(p_0)G_N(E_A+\nu-p_0,\bmq) \ ,
\label{a4}
\end{eqnarray}
and where we shall use the spectral representation of $G_{A-1}$
\begin{equation}
G_{A-1}(p_0)=
\sum_n{|\Phi_{A-1}^{(n)}\rangle
\langle \Phi_{A-1}^{(n)}|\over 
p_0-E_{A-1}^{(n)} +i\eta}\ .
\label{a5}
\end{equation}
$G_N$ in Eq.~(\ref{a4}) stands for the Green's function
of the struck nucleon after absorption of the virtual photon. It reads  
\begin{equation}
G_N(E_A+\nu -p_0,\bmq ) =
{1\over E_A+\nu -p_0-(\hat\bmp +\bmq )^2/2M  +i\eta}\ .
\label{a6}
\end{equation}

Substituting Eq.~(\ref{a4}) into Eq.~(\ref{a1})
and performing the integration over $p_0$, one obtains the structure 
function in the IA 
\begin{mathletters}
\label{a7}
\begin{eqnarray}
W^{IA}(\nu,\bmq )&=&\sum_n\int \frac {d\bmp}{(2\pi)^3}|\varphi_A^{(n)}(\bmp )|^2
\delta \left(E_A +\nu -E_{A-1}^{(n)}- {(\bmp -\bmq)^2\over 2M}\right )
\label{a7a}\\
&=&\int \frac {d\bmp}{(2\pi)^3}\int dE {\cal P}(\bmp,E)
\delta \bigg (\nu-E-\Delta-\frac{(\bmp-\bmq)^2}{2M}\bigg )\ ,
\label{a7b}
\end{eqnarray}
\end{mathletters}
where  $\varphi_A^{(n)}(\bmp )=\langle\Phi_{A-1}^{(n)},\bmp
|\Phi_A\rangle$ is an overlap amplitude and 
$\Delta=E_{A-1}-E_A$. We neglect in  $E_{A-1}^{(n)}$  the tiny 
recoil-energy of the spectator $p^2/2M_{A-1}$.
In Eq. (\ref{a7b}) appears the single-hole spectral function
\begin{equation}         
{\cal P}(\bmp,E)=\sum_n |\varphi_A^{(n)}(\bmp )|^2
\delta(E-{\cal E}_n), 
\label{a8}
\end{equation}
with ${\cal E}_n\equiv E_{A-1}^{(n)}-E_{A-1}$, the spectator excitation 
energy.  

It will be useful to define the reduced structure function
for non-relativistic systems
\begin{equation}
F(y,\bmq )=(\qq/M)W(\nu,\bmq )\ ,
\label{a9}
\end{equation}
where $y\equiv y(\nu,\bmq )$ is some scaling variable.
After integration in Eq. (\ref{a7b}) over $\hat \bmp.\hat \bmq$ 
one obtains for the lowest order, IA part of $F$
\begin{eqnarray}
F^{IA}(y_0,\bmq )=\frac {1}{4\pi^2}\bigg \lbrack
\int_{|y_0|}^{y_0+2q} dpp\int_0^{E_{max}}dE {\cal P}(\bmp,E)
+\theta(y_0) \int_0^{y_0} dpp\int_{E_{min}}^{E_{max}}dE
{\cal P}(\bmp,E) \bigg \rbrack,
\label{a10}
\end{eqnarray}
with $y_0$, the IA scaling variable
\begin{eqnarray} 
y_0=-\qq+\sqrt {2M(\nu-\Delta)} \ .
\label{a11}
\end{eqnarray}
The integration limits in Eq. (\ref{a10}) are
\begin{eqnarray} 
E_{max\atop min}(y_0,p,q)={y_0\pm p\over M}\qq 
+{y_0^2-p^2\over 2M}\ , 
\label{a12}
\end{eqnarray}
and in particular
\begin{equation}
\lim_{q\to\infty}\ E_{max}(y_0,p,q)=\infty \ .
\label{a13}
\end{equation}

In order to go beyond the IA one expands the total
Green's function $G_{1,A}$, Eq.~(\ref{a2}), 
in powers of $V_1G_{1,A}^{(0)}$. Substituting this expansion 
into Eq.~(\ref{a1}) one obtains  the  Impulse Series  (IS) 
for  the structure function  
\begin{equation}
W=-\frac{1}{\pi}{\rm Im}
\langle \Phi_A|G_{1,A}^{(0)}+G_{1,A}^{(0)}V_1G_{1,A}^{(0)}
+G_{1,A}^{(0)}V_1G_{1,A}^{(0)}V_1G_{1,A}^{(0)} \ .
+\cdots |\Phi_A\rangle \ .
\label{a14}
\end{equation}
The first term is the IA  and the remainder are FSI.  We now introduce 
the scattering operator $T$, which  describes 
the scattering of the knocked-out nucleon from the ($A-1$)-nucleon spectator. 
It satisfies the Lippmann-Schwinger operator equation 
$T=V_1+V_1G_{1,A}^{(0)}T$, and clearly permits a formal 
summation of the FSI terms in Eq. (\ref{a14}). The 
total structure function thus becomes
\begin{eqnarray}
W=-\frac{1}{\pi}{\rm Im}
\langle \Phi_A|G_{1,A}^{(0)}+G_{1,A}^{(0)}TG_{1,A}^{(0)}| \Phi_A\rangle \ .
\label{a15}
\end{eqnarray}
It is convenient to use the momentum representation for the nucleon and, 
as in Eqs. (\ref{a5}), (\ref{a7}),
a representation for the spectator states, denoted by $'n'$.
The Lippmann-Schwinger equation then becomes a set of coupled equations
for transition amplitudes $T_{nn'}(E,\bmp,\bmp') \equiv 
\langle \bmp,\Phi_{A-1}^{(n)}|T|\Phi_{A-1}^{(n')},\bmp'\rangle$

\begin{eqnarray}
T_{nn'}(E, \bmp,\bmp')=
V_{1;nn'}(\bmp -\bmp' ) +\sum_{n''}\int {d\bmp''\over
(2\pi)^3} 
{V_{1;nn''}(\bmp -\bmp'' )T_{n''n'}(E,\bmp'',\bmp')
\over E-{\cal E}_{n''}-{{\bmp''}^2\over 2M} +i\eta} ,
\label{a16}
\end{eqnarray}
Here $V_{1;nn'}(\bmp -\bmp' )=\langle \bmp ,\Phi_{A-1}^{(n)}|V_1|
\Phi_{A-1}^{(n')}, \bmp'\rangle$ and $E$, the  energy in the lab frame.
In parallel the  total reduced response $F(y_0,\bmq )$, Eq. (\ref{a9}), 
reads  (we chose the $z$-axis along $\bmq$)
\begin{eqnarray}
F(y_0,\bmq )&&=F^{IA}(y_0,\bmq )+{M\over \pi\qq }
{\mbox{Im}} \sum_{nn'}\int\frac {d\bmp d\bmp'}{(2\pi)^6}
\nonumber\\*[5pt]
&&\frac {\varphi_A^{(n)}(\bmp )T_{nn'}(E_{N,A-1},\bmp +\bmq ,\bmp' +\bmq )
\varphi_A^{(n')}(\bmp' )}
{\left (y_0-p_z-{\displaystyle M{\cal E}_n\over\qq}-
{\displaystyle \bmp^2-y_0^2\over
\displaystyle 2\qq} +i\eta\right ) 
\left (y_0-p'_z-{\displaystyle M{\cal E}_{n'}\over\qq}-
{\displaystyle {\bmp'}^2-y_0^2\over 
\displaystyle 2\qq}+i\eta\right )},  
\label{a17} 
\end{eqnarray}
with
\begin{equation}
E_{N,A-1}=\nu -\Delta = {(y_0+\qq )^2\over 2M} \ ,
\label{a18}
\end{equation}
the off-shell energy of the nucleon-spectator  amplitudes.

\subsection{The GRS series.}

The expansion  (\ref{a14}) of a structure function 
in powers of the residual interaction $V_1$ 
is not the only possible perturbative  approach.  
In this section we shall expand  
the  shifted Green's function $G_{1,A}(E_A+\nu,\bmq )$ 
in  a different operator 
$\tilde V=V_1+K_1(0)+H_{A-1}-E_A\equiv -G_{1,A}^{-1}(E_A,0)$,  for which by
definition $\tilde V|\Phi_A\rangle=0 $. Then using the identity 
\begin{equation}
G_{1,A}(E_A+\nu,\bmq )={1\over G_{1,A}^{-1}(E_A+\nu,\bmq )-
G_{1,A}^{-1}(E_A,0)-\tilde V} \ ,
\label{a19}
\end{equation}
the shifted Green's function $G_{1,A}(E_A+\nu,\bmq )$ permits the expansion  
\begin{eqnarray}
G_{1,A}=\tilde G_{1,A}(1+\tilde V\tilde G_{1,A}+
\tilde V\tilde G_{1,A} \tilde V\tilde G_{1,A}
+\cdots )\ ,
\label{a20}
\end{eqnarray}
where
\begin{mathletters}
\label{a21}
\begin{eqnarray}
\tilde G_{1,A}\equiv \tilde G_{1,A}(\nu ,\bmq )&=&
{1\over G_{1,A}^{-1}(E_A+\nu ,\bmq )-G_{1,A}^{-1}(E_A,0 )}
\label{a21a}
\\*[5pt]
&=&{1\over [G_{1,A}^{(0)}(E_A+\nu ,\bmq )]^{-1}-[G_{1,A}^{(0)}(E_A,0 )]^{-1}}\ .
\label{a21b}
\end{eqnarray}
\end{mathletters}
Again we assume  $V_1$ to be local and it therefore cancels out in
Eq.~(\ref{a21b}).
Expressing  $\tilde G_{1,A}$ as a convolution (cf. Eq.~(\ref{a4}), 
Eq.~(\ref{a21b}) becomes
\begin{eqnarray}
\tilde G_{1,A}(\nu ,\bmq )&=&i\int {dp_0\over 2\pi}
G_{A-1}(p_0)
{1\over G^{-1}_N(E_A+\nu -p_0 ,\bmq )-G^{-1}_N(E_A-p_0,0 )}
\nonumber\\*[5pt]
&=&i\int {dp_0\over 2\pi} G_{A-1}(p_0)\tilde G_N(\nu, \bmq )
\ ,
\label{a22}
\end{eqnarray}
Since $G_N^{-1}$, Eq.~(\ref{a6}), is linear in the energy argument, 
the spectator  energy $p_0$ and $E_A$ cancel in the denominator in
Eq.~(\ref{a22}). Thus in contrast to $G_{1,A}^{(0)}$,
Eqs.~(\ref{a4})-(\ref{a6}), $\tilde G_{1,A}$ does not 
depend on  
the excitation energy $E_{A-1}^{(n)}$ of the spectator. 
Using Eq.~(\ref{a5}), one performs the $p_0$ integral in (\ref{a22}) 
with the result 
\begin{eqnarray}
\tilde G_{1,A}(\nu ,\bmq )\equiv \tilde G_N(\nu ,\bmq )
={M\over \qq}
{1\over y_W-\hat p_z+i\eta}\ ,
\label{a23}
\end{eqnarray}
where $y_W$ is the GRS-West scaling variable \cite{grs,west}
\begin{eqnarray}
y_W={M \over \qq}\bigg (\nu-\frac {q^2}{2M}\bigg )\ .
\label{a24}
\end{eqnarray}

Substitution of the series (\ref{a20}) for $G_{1,A}$ into Eq. (\ref{a1}), and
use of Eq. (\ref{a23}) there, manifestly produces a power series in 
$\tilde V/\qq$ (the GRS series) for the nuclear response 
\begin{mathletters}
\label{a25}
\begin{eqnarray}
W(\nu ,\bmq )&=&-\frac{1}{\pi}{\rm Im}
\langle \Phi_A|\tilde G_N+\tilde G_N\tilde V\tilde G_N
+\tilde G_N\tilde V\tilde G_N\tilde V\tilde G_N
\cdots |\Phi_A\rangle
\label{a25a}\\*[5pt] 
&=&\sum_{j=0}^{\infty} \left (\frac{M}{\qq}\right )^{j+1} F_j(y_W) \ ,
\label{a25b}
\end{eqnarray}
\end{mathletters}
with coefficients $F_j$, which  are functions of  the scaling variable 
$y_W$. The lowest order  GRS term ($j=0$) is
the asymptotic limit $\bmq\to\infty$, of the reduced structure function
Eq. (\ref{a9}),  
\begin{eqnarray}
F^{GRS}_0(y_W)=\int n(p)\delta(y_W-p_z)\frac{d^3p}{(2\pi)^3}
=\frac{1}{4\pi^2}\int_{|y_W|}^{\infty}n(p)p\ dp\ . 
\label{a26}
\end{eqnarray}
Above $n(p)$ is the nucleon momentum distribution, which is related to
the spectral function Eq.~(\ref{a8}) by
\begin{equation}
n(p)=\int_{0}^\infty {\cal P}(p,E)dE. 
\label{a27}
\end{equation}
We remark that the leading terms in the Impulse and GRS series,
Eqs. (\ref{a10}) and  (\ref{a26}) are quite different.
However, using  $\lim_{\qq\to\infty}(y_W-y_0)=0 $ and 
Eqs. (\ref{a13}) and (\ref{a27}), one finds that in the limit
$|\bmq|\to\infty$, $F^{IA}\to F_0^{GRS}$.

Consider next higher order terms 
$\langle \Phi_A|\tilde G_N(\tilde V\tilde G_N)^n|\Phi_A\rangle$
in the series (\ref{a25}). 
Since  $[\tilde V,\tilde G_N]=[V_1,\tilde G_N]$ and  also 
$\tilde V|\Phi_A\rangle=0$, each of those terms can be expressed by
commutators, involving the residual interaction $V_1$ and the kinetic
energy operator $K_1$ of the struck nucleon, and  
not $\tilde V=H_A-E_A$. For instance 
\begin{eqnarray} 
&&\tilde V\tilde G_N|\Phi_A\rangle =[V_1,G_N^{(0)}]|\Phi_A\rangle \ ,
\nonumber\\*[5pt]
&&\tilde V\tilde G_N\tilde V\tilde G_N|\Phi_A\rangle =
\{[V_1,\tilde G_N]^2+[(V_1+K_1),[V_1,\tilde G_N]]\}|\Phi_A\rangle \
\label{a28}\\*[5pt]
&&~~~~~~~~~~~~~~~~~~~~~~~~~~~~~~\cdots
\nonumber
\end{eqnarray}
From Eq. (\ref{a25}) one  then finds for the 
corresponding reduced structure function Eq. (\ref{a9})  
\begin{eqnarray}
F^{GRS}=-\frac{\qq}{\pi M }{\rm Im}
\langle \Phi_A|\tilde G_N+[\tilde G_N,V_1]\tilde G_N+
[\tilde G_N,V_1]\tilde G_N[V_1,\tilde G_N]+
\cdots |\Phi_A\rangle \ .
\label{a29}
\end{eqnarray} 
Eq.~(\ref{a29}) is the GRS series 
for the response function which, using a coordinate-time representation, 
has first been derived in  Ref. \cite{grs}.
For instance, the leading  FSI term $F_1(y_W)$ reads
\begin{mathletters}
\label{a30} 
\begin{eqnarray}
&&F_1^{GRS}(y_W)= 
\frac{1}{\pi}{\rm Im}\sum_{nn'}\int \frac {d\bmp\ d\bmp'}{(2\pi)^6}
\frac {\varphi_A^{(n)}(\bmp)
V_{1,nn'}(\bmp-\bmp')(p_z'-p_z)\varphi_A^{(n')}(\bmp')}
{(y_W-p_z+i\eta)(y_W-p_z'+i\eta)^2}
\label{a30a}\\
&&=-i\int\limits_{-\infty}^{\infty}\frac {ds}{2\pi}e^{isy_W}
\int \int d\bmr_1d\bmr_2
\rho_2(\bmr_1-s\hat\bmq,\bmr_2;\bmr_1,\bmr_2)
\int_0^s d\sigma
[V_{12}(\bmr-\sigma \hat \bmq)-V_{12}(\bmr-s \hat \bmq)] \ ,
\label{a30b}
\end{eqnarray}
\end{mathletters}
with $\rho_2$ is the 2-particle density matrix.

In spite of the increasing complexity of the commutators in the series 
(\ref{a29}), it has been demonstrated in Ref.\cite{sag1} that, like
Eq.~(\ref{a15}) for the IS, $all$ FSI
terms in  the GRS series, Eq.~(\ref{a29}), can be summed in a closed 
expression.
\begin{eqnarray}
F=-\frac{\qq}{\pi M}{\rm Im}
\langle \Phi_A|\tilde G_N+G_{1,A}^{(0)}\tilde G_N^{-1}
[\tilde G_N,T]G_{1,A}^{(0)}|\Phi_A\rangle\ .
\label{a31}
\end{eqnarray}
with  $G_{1,A}^{(0)}$, given by Eq.~(\ref{a3}). A derivation of Eq.~(\ref{a31}) 
is given in the Appendix. 

The characteristic feature of the expression  (\ref{a31}) is the commutator 
$[\tilde G_N,T]$, involving $T$, Eq.~(\ref{a16}), which describes the 
scattering of the struck nucleon and the spectator. That commutator 
has a simple form in the momentum representation
\begin{eqnarray}
\langle \bmp |\tilde G_N^{-1}[\tilde G_N,T]|\bmp' \rangle =
\langle \bmp |T|\bmp' \rangle{p_z-p'_z\over y_W-p_z'+i\eta} \ .
\label{a32}
\end{eqnarray}
Using the spectral representation of the
Green's function  $G_{1,A}^{(0)}$, Eqs.~(\ref{a4})-(\ref{a5})
one rewrites Eq.~(\ref{a31}) as 
\begin{eqnarray}
F(y_W,\bmq )&&=F_0^{GRS}(y_W)+{M\over \pi\qq }
{\mbox{Im}} \sum_{nn'}\int\frac {d\bmp d\bmp'}{(2\pi)^6}\nonumber\\*[5pt]
&&\frac {\varphi_A^{(n)}(\bmp )(p'_z-p_z)T_{nn'}
(\tilde E_{N,A-1};\bmp +\bmq ,\bmp' +\bmq )
\varphi_A^{(n')}(\bmp' )}
{\left (y_W-p_z-{M\Delta_n(p)\over\qq}  +i\eta\right )
\left (y_W-p'_z-{M\Delta_{n'}(p')\over\qq} +i\eta\right )
(y_W-p'_z+i\eta)} \ , 
\label{a33}
\end{eqnarray}
where
\begin{equation}
\Delta_n(p)=\Delta +{\cal E}_n+{p^2\over 2M}\ ,
\label{a34}
\end{equation}
and
\begin{equation}
\tilde E_{N,A-1}=\nu-\Delta ={(y_W+\qq )^2\over 2M}-\Delta -{p^2\over 2M}
\label{a35}
\end{equation}
is the off-shell energy of $T$ in the lab frame.

Expansion of  the integrand (\ref{a33}) in powers of $1/\qq$ for constant  
$y_W$ generates the entire GRS series, Eq.~(\ref{a29}).
For instance, the leading  FSI term of the GRS series $F_1^{GRS}(y_W )$,
Eq.~(\ref{a30}),  is retrieved  from Eq.~(\ref{a33}) by the replacement 
$T\to V_1$ and disregarding 
$M\Delta_n(p)/\qq$. Likewise one assembles
terms of higher order in $1/\qq$, all appearing as sums over $n$. Those may in
fact be evaluated and  ultimately produce, as in 
the original presentation of the GRS theory, 
coefficients $F_j$ in terms of
off-diagonal density matrices \cite{grs} (cf. Eq.~(\ref{a30b}) for $F_1$). 

The  expressions  Eqs.~(\ref{a17}), (\ref{a33}) permit a comparison of
the total FSI contributions in the IS and GRS series.
Both contain nucleon-spectator transition amplitudes, which are strongly peaked  
for small momentum transfers $p'_z-p_z$. However, the same momentum transfer 
also  appears as a factor in the numerator of Eq.~(\ref{a33}) and thus 
reduces FSI in the GRS series.

An additional suppression of FSI in that series comes from the
different off-shell energies Eqs. (\ref{a18}) and ({\ref{a35}). From those one
finds for $y_0=y_W$, $\tilde E_{N,A-1}<E_{N,A-1}$, i.e. the energy of the
GRS amplitude is farther from the energy shell than is IS amplitude.
Since the complete expressions for the structure functions 
are identical, the forwarded arguments  indicate that the  leading GRS term 
$F^{GRS}_0$ is a  better approximation to the total structure function
than is the corresponding $F^{IA}$.
Experimental evidence is deferred
to the end of Section IV.

\section{Relativistic nuclear structure function}

In Section II we have used an unconventional propagator technique 
to re-derive the
GRS series, primarily because the same will now be shown to lead to the
desired relativistic generalization of the GRS series, Eq.~(\ref{a29}).
 
We start with the relativistic nuclear structure function
$W_{\mu \nu}$. As in the previous case we consider for 
simplicity scalar nucleons and photons. This 
implies that we restrict ourself to the longitudinal component of the
structure function $W=[(\bmq^2-\nu^2)/\bmq^2]W_{00}$ (see for
instance\cite{west,mul}). We presume that the techniques which we shall
present below, will also be applicable for nucleons and photons with spin.   

The relativistic nuclear structure function is then again given by 
the imaginary part of the forward Compton amplitude.  
The latter can always be written as a sum of 
two terms, which represent the IA and FSI contributions  
(Fig.~1)
\begin{eqnarray}
W(q)&=&-\frac{1}{\pi}{\rm Im} \left \{\Gamma_A G_N(P_A)\left
[G_{1,A}^{(0)}(P_A+q)\right.\right.\nonumber\\*[5pt]
&&\left. \left. + G_{1,A}^{(0)}(P_A+q)T(P_A+q)G_{1,A}^{(0)}(P_A+q)\right ]
G_N(P_A)\Gamma_A\right \} .   
\label{a36}
\end{eqnarray}
$G_N$ and $G_{1,A}^{(0)}$ are propagators for, respectively,  a nucleon and
the non-interacting  nucleon-spectator system, with 4-momentum
$P_A+q$. As before we display   in  Eq.~(\ref{a36}) only
the external parameters $P_A=(M_A,0)$ and $q=(\nu ,\bmq )$. 
Only when necessary, do we make explicit  the 4-momenta  of target nucleons. 
Those appear for example in $G_N$
\begin{equation}
G_N(P_A)\equiv G_N(P_A-p)={1\over (P_A-p)^2-M^2+i\eta}
\label{a37}
\end{equation}
and likewise in $G_{1,A}^{(0)}$ 
\begin{equation}
G_{1,A}^{(0)}(P_A+q)\equiv G_{1,A}^{(0)}(P_A+q,p)=iG_{A-1}(p)G_N(P_A+q-p),
\label{a38}
\end{equation}
where $G_{A-1}$ is the propagator of the fully interacting spectator.
The operator $T=T(P_A+q)$ in Eq.~(\ref{a36}) again describes elastic and
inelastic scattering of the $N$-spectator sub-systems and satisfies 
the Bethe-Salpeter equation  (cf. Eq. (\ref{a16}))
\begin{eqnarray}
T(P_A+q)=V_1(P_A+q)[1+G_{1,A}^{(0)}(P_A+q)T(P_A+q)] \ .
\label{a39}
\end{eqnarray}
The  effective interaction $V_1$ is defined as 
the sum of all irreducible contributions, 
which drive the scattering operator in Eq. (\ref{a39}).
\vskip1cm
%%%%%%%%%%%%%%%%%%%%%%%%%%%%%%%%%%%%%%%%%%%%%%%%%%%%%%%%%%%%%%%%%%%%%%%%%
\begin{minipage}{13cm}
\begin{center}
\leavevmode
\epsfxsize=15cm
\epsffile{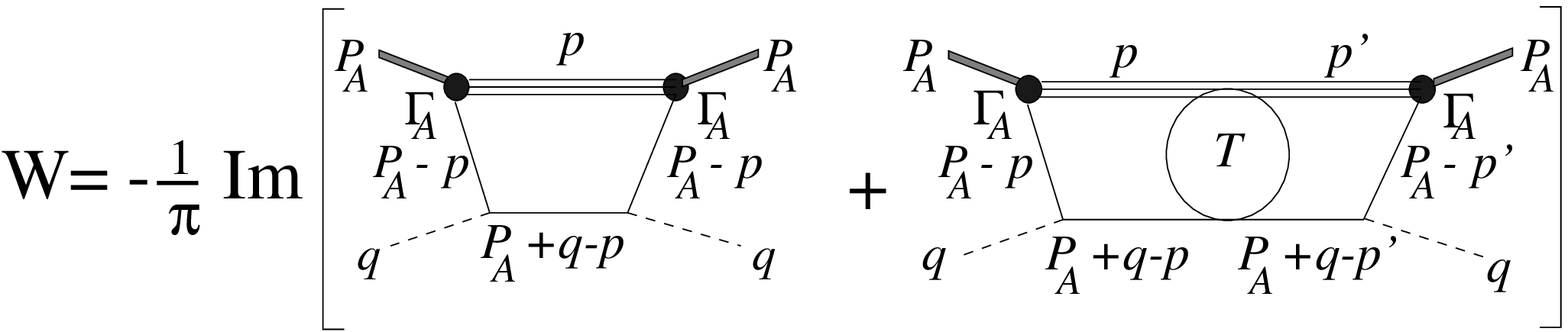}
\end{center}
{\begin{small}
Fig.~1. Nuclear structure function expressed
as the imaginary part of the forward Compton amplitude.
The first diagram represents the IA and the second one  FSI.
\end{small}}
\end{minipage} \\ \\
%%%%%%%%%%%%%%%%%%%%%%%%%%%%%%%%%%%%%%%%%%%%%%%%%%%%%%%%%%%%%%%%%%%%%%%%%

We still have to define the  target-spectator-$N$ vertex 
function $\Gamma_A$ in Eq.~(\ref{a36}) (see also
Fig.~1). It appears in the residue of the bound state pole of the 
scattering operator $T(P)$
\begin{eqnarray}
\Gamma_A(p)\Gamma_A(p')={\lim_{P^2\to M_A^2}} (P^2-M_A^2)
\langle p|T(P)|p'\rangle \ .
\label{a40}
\end{eqnarray}
One then derives from the Bethe-Salpeter equation 
(\ref{a39}) with  the  4-momentum  $P_A$ as argument  
\begin{eqnarray}
\Gamma_A(p)=i\int\langle p|V_1(P_A)|p'\rangle G_{A-1}(p')
G_N(P_A-p')\Gamma_A(p'){d^4p'\over (2\pi)^4}\ ,
\label{a41}
\end{eqnarray}
which is the Dyson equation, satisfied by $\Gamma_A$. The latter can be 
rewritten in a form, similar to the Schr\"odinger equation 
\begin{equation}
\left \{[G_{1,A}^{(0)}(P_A)]^{-1}-V_1(P_A)\right \}G_{1,A}^{(0)}(P_A)
\Gamma_A=0\ ,
\label{a42}
\end{equation}
with $G_{1,A}^{(0)}(P_A)\Gamma_A$  a relativistic target wave function.

Next we link in a standard way the Green's function of the fully
interacting $A$-nucleon target with $G_{1,A}^{(0)}$ and the scattering
operator (cf. Eqs.~(\ref{a2}), (\ref{a15}))
\begin{eqnarray}
G_{1,A}(P_A+q)&=& G_{1,A}^{(0)}(P_A+q)[1+ T(P_A+q)G_{1,A}^{(0)}(P_A+q)]
\nonumber\\
&=&{1\over [G_{1,A}^{(0)}(P_A+q)]^{-1}-V_1(P_A+q)} \ .
\label{a43}
\end{eqnarray}
The momentum $q$ of the virtual photon in the
argument of the total Green's function is ultimately the one, absorbed by
nucleon $'1'$. Eq.~(\ref{a36}) can then  be rewritten as
\begin{eqnarray}
W(q)=-\frac{1}{\pi}{\rm Im} [\Gamma_A G_N(P_A)G_{1,A}(P_A+q)G_N(P_A)\Gamma_A], 
\label{a44} 
\end{eqnarray}
Eqs.~(\ref{a43}), (\ref{a44}) are the 
relativistic analogs of Eqs.~(\ref{a1}), (\ref{a2}). Whereas  the
latter have been derived by explicit use of a Hamiltonian, this is not 
so for the former.

The above equations  serve as 
the starting point for various perturbative approaches for 
the structure function. First one expands $G_{1,A}(P_A+q)$ in 
powers of $V_1$ which produces the 4-dimensional relativistic IS (cf. 
(\ref{a14}))   
\begin{eqnarray}
W(q)&=&-\frac{1}{\pi}{\rm Im} \left \{\Gamma_A G_N(P_A)\left 
[G_{1,A}^{(0)}(P_A+q)
\right.\right.\nonumber\\
&+&\left.\left.
G_{1,A}^{(0)}(P_A+q)V_1(P_A+q)G_{1,A}^{(0)}(P_A+q)+\cdots\right ]
G_N(P_A)\Gamma_A\right \} \ . 
\label{a45}
\end{eqnarray}

As for the non-relativistic case (see paragraph before Eq. (\ref{a28}))
we next look for a different expansion of $W$ in powers of an operator 
$\tilde V$ which annihilates the target ground state. A choice which 
satisfies this requirement is provided by  the bracketed operator in 
Eq.~(\ref{a42}) 
\begin{equation}
\tilde V(P_A)=V_1(P_A)-[G_{1,A}^{(0)}(P_A)]^{-1} \ .
\label{a46}
\end{equation}
Using  Eq. (\ref{a46}) we then rewrite $G_{1,A}(P_A+q)$, Eq.~(\ref{a43}), as
\begin{equation}
G_{1,A}(P_A+q)={1\over [G_{1,A}^{(0)}(P_A+q)]^{-1}-[G_{1,A}^{(0)}(P_A)]^{-1}
-V_1(P_A+q)+V_1(P_A)-\tilde V(P_A)} \ .
\label{a47}
\end{equation}
For further evaluation we assume that the interaction between the 
$N$ and the spectator is the sum of local pair potentials, each 
depending only on the 4-momentum transfer 
\begin{eqnarray}
\langle p_1,p_2,...,p_k,..|V_1|p'_1,p'_2,....,p_k',...\rangle=\sum_{k\ge 2} 
V_{1k}(p_1-p_1')
\delta^{(4)}(p_1-p_k-p'_1+p_k') \ .
\label{a48}
\end{eqnarray}
As a consequence $V_1(P_A+q)-V_1(P_A)=0$, in (\ref{a47}).
Expanding there $G_{1,A}(P_A+q)$ in powers of $\tilde V$ and substituting 
the result into Eq.~(\ref{a44}), one obtains (cf. Eq. (\ref{a25a}))
\begin{eqnarray}
W(q)&=&-\frac{1}{\pi}{\rm Im} \left \{\Gamma_A G_N(P_A)\left
[\tilde G_{1,A}(P_A,q) \right.\right.\nonumber\\
&+&\left.\left.
\tilde G_{1,A}(P_A,q)\tilde V(P_A)\tilde G_{1,A}(P_A,q)+\cdots\right ]
G_N(P_A)\Gamma_A\right \} \ , 
\label{a49}
\end{eqnarray}
with 
\begin{eqnarray}
\tilde G_{1,A}(P_A,q)&=&{1\over
[G_{1,A}^{(0)}(P_A+q)]^{-1}-[G_{1,A}^{(0)}(P_A)]^{-1}}\nonumber\\[5pt]
&=&iG_{A-1}(p)
{1\over G_N^{-1}(P_A+q-p)-G_N^{-1}(P_A-p)}\equiv G_{A-1}(p)
\tilde G_N(P_A-p,q)
\ . 
\label{a50}
\end{eqnarray}
For clarity we made explicit the 4-momentum of the struck nucleon. 

We now evaluate the modified  Green's function of the struck nucleon, 
$\tilde G_N$ in Eq. (\ref{a50}). Using  Eq.~(\ref{a37}) one obtains
\begin{eqnarray}
\tilde G_N(P_A-p,q)
&=&\frac{1}{(P_A-p+q)^2-(P_A-p)^2+i\eta}\nonumber\\
\noalign{\vskip5pt}
&=&\frac{1}{2(M_A-p_0)\nu-2p_z\qq-Q^2+i\eta} \ ,
\label{a51}
\end{eqnarray}
with $Q^2=\bmq^2-\nu^2$, and where the negative $z$ axis has been chosen 
in the direction of the momentum the virtual photon. One notes 
that in contrast to the 
non-relativistic case Eqs.~(\ref{a22}), (\ref{a23}), the quadratic 
dependence on energy in the relativistic propagator   Eq.~(\ref{a37}) 
causes the spectator energy $p_0$  to persist in  Eq.~(\ref{a51}). 

Next one  exploits Eq.~(\ref{a42}) in  order to replace $\tilde V$ 
in each term of this series  by commutators involving 
the residual interaction, $V_1$ \cite{sag1,sag2}. For instance,
the leading FSI term (the second term of the expansion (\ref{a49}))
becomes
\begin{eqnarray}
W_1^{GRS}(q)&=&-\frac{1}{\pi}{\rm Im} \int {d^4pd^4p'\over (2\pi )^8}
\Gamma_A(p) G_N(P_A-p)G_{A-1}(p)
[\tilde G_N(P_A-p,q),V_1(p-p')]\nonumber\\*[5pt]
&&\tilde G_N(P_A-p',q)G_{A-1}(p')G_N(P_A-p')\Gamma_A(p') \ ,
\label{a52}
\end{eqnarray} 
where we made explicit the momentum of the struck nucleon, but left implicit
variables chosen to represent the spectator  nucleons. 
The  entire series  formally acquires the same form as its non-relativistic 
GRS counterpart, Eq.~(\ref{a29}), but each term contains 4-dimensional integrals 
over intermediate 4-momenta. 

The exact evaluation of these terms, as well of those in the relativistic
IS,  constitutes a formidable many-body problem. We now
discuss minimal assumptions which lead to considerable simplifications.  

\section{3-dimensional reduction}

\subsection{Non-relativistic limit for target wave functions.}

We start this section with the observation that nucleons in ground
states of nuclei and in not 
too highly excited states have on the average 3-momenta 
$\langle\bmp^2\rangle^{1/2} \lesssim p_F\approx$ 0.3 GeV, with $p_F$
the Fermi momentum.
The above are thus  essentially non-relativistic systems. Examples are 
the struck nucleon before the absorption of the virtual photon,  the 
nucleons in the target nucleus at rest and in the spectator, which recoils 
with  momentum $\bmp$. Only particles or sub-systems with momenta
containing  $q$ are truly relativistic. As Fig.~1 shows, this applies 
only to the recoiling nucleon with momentum $p+q\approx q$.

We  thus apply  non-relativistic limits to all quantities  which
contain  low-momentum nucleons. Those
are the propagators $G_N(P_A-p)$ and  $G_{A-1}(p)$, Eq.~(\ref{a37}),
(\ref{a38}) (cf. Eqs.~(\ref{a5}), (\ref{a6}))
\begin{equation}
G_N(P_A-p) \simeq \left ({\displaystyle 1\over \displaystyle 2M}\right )
{\displaystyle 1  \over\displaystyle M_A-p_0-M-\bmp^2 /2M+i\eta}
\label{a53} 
\end{equation}
and 
\begin{equation}
G_{A-1}(p)\simeq \sum_n\left ({\displaystyle 1
\over \displaystyle 2M_{A-1}}\right )
{\displaystyle |\Phi_{A-1}^{(n)},\bmp\rangle
\langle\bmp ,\Phi_{A-1}^{(n)}|\over\displaystyle
p_0-M_{A-1}-{\cal E}_n+i\eta}\ . 
\label{a54} 
\end{equation}
In the same limit one can use for the residual interaction $V_1(p-p')\approx V_1(\bmp-\bmp')$
and for the vertex function 
$\Gamma_A(p)\approx \Gamma_A(\bmp)$.
After substitution of the above limits into Eq.~(\ref{a41}) we consider
the integration over $p_0$. One notes that the Green's functions $G_N(P_A-p)$ 
and  $G_{A-1}(p)$ have poles in the complex $p_0$-plane which lie on different 
sides of the real axis.  One may thus close  the integration contour
around the spectator pole and perform the $p_0$ integration. The result is
\begin{equation}
\left (E_A-H_{A-1}-K_1-{1\over 4M_{A-1}M}V_1\right )
\Phi_A=0
\label{a55}
\end{equation}
with 
\begin{equation}
\Phi_A={1\over (8M_{A-1}M^2)^{1/2}(E_A-H_{A-1}-K_1)}\Gamma_A  \ . 
\label{a56}
\end{equation}
Eq.~(\ref{a56}) is now a standard 3-dimensional Schr\"odinger equation for 
the  target bound state wave function, with effective residual 
interaction $(1/4M_{A-1}M)V_1$. 

\subsection{Reduction of Relativistic Impulse Series} 

We consider the relativistic IS and first 
apply  the above non-relativistic limits to all quantities,
depending on nucleons with low momenta. This we illustrate below on 
the IA for the structure function, 
$W^{IA}=-(1/\pi) {\rm Im}[\Gamma_A G_N G^{(0)}_{1,A}G_N \Gamma_A]$,
which
is  the first term of the IS, Eq. (\ref{a45}). Explicitly
\begin{eqnarray}
W^{IA}(\nu,\bmq )&=&-\frac{1}{\pi}{\rm Im} \sum_n
\int \frac {d^3p}{(2\pi)^3} \int \frac {dp_0}{2\pi} {i(8M_{A-1}M^2)^{-1}
|\langle\Gamma_A(\bmp )|\Phi_{A-1}^{(n)}\rangle |^2\over
\left (M_A-p_0-M-{\bmp^2\over 2M}+i\eta\right)^2
(p_0-M_{A-1}-{\cal E}_n+i\eta)}
\nonumber\\*[5pt]
&&{1\over (M_A+\nu -p_0)^2-e^2_{\bmq-\bmp}+i\eta} \ ,
\label{a57}
\end{eqnarray}
with $e_{\bmp}=\sqrt{M^2+\bmp^2}$. 

One  observes that the above-mentioned spectator pole, 
$p_0=M_{A-1}+{\cal E}_n-i\eta$  and  the negative  energy nucleon pole in 
the relativistic propagator $G_N(P_A+q)$ at 
$p_0=M_A+\nu +e_{\bmq -\bmp}-i\eta$ lie both in the lower 
half of the complex $p_0$-plane. One ought to include the two 
above-mentioned poles, but we first disregard the one with negative-energy
and compute only the residue  of the spectator pole 
leading to
\begin{eqnarray}
W^{IA}(\nu,\bmq )=\sum_n\int 
{d^3p\over  2e_{\bmq -\bmp}(2\pi )^3}|\varphi_A^{(n)}(\bmp )|^2\delta \left
(\nu -\Delta -{\cal E}_n+M-e_{\bmq -\bmp)} \right ) \ .  
\label{a58}
\end{eqnarray}
Next we introduce the reduced relativistic structure function
\begin{eqnarray}
{\cal F}(\bar y_0,\bmq )\equiv2\qq W(\nu ,\bmq),
\label{a59}
\end{eqnarray}
where the factor $2\qq$ has been adjusted to produce the correct
non-relativistic limit Eq. (\ref{a9}) of ${\cal F}$.
Integration over $\cos (\bmp ,\bmq )$ leads to 
\begin{eqnarray}
{\cal F}^{IA}(\bar y_0,\bmq )={1\over 4\pi^2}\left [
\int\limits_{|\bar y_0|}^{2q+\bar y_0}pdp
\int\limits_0^{\bar E_{max}}  {\cal P}(p,E)dE
+\theta (\bar y_0)\int\limits_0^{\bar y_0}pdp\int
\limits_{\bar E_{min}}^{\bar E_{max}}
{\cal P}(p,E)dE\right ]\ .   
\label{a60}
\end{eqnarray}
It has the same form as the non-relativistic IA, Eqs. (\ref{a10}), where the 
scaling variable and the integration limits have been replaced by relativistic
ones 
\begin{eqnarray}
&&\bar y_0=-\qq +\sqrt{2M(\nu -\Delta )+(\nu -\Delta )^2}
\label{a61}\\*[5pt]
&&\bar E_{max\atop min}(q ,\bar y_0,p)=e_{\displaystyle \bar y_0+\qq }
-e_{\displaystyle p\pm\qq } \ .   
\label{a62}
\end{eqnarray}
In contrast with Eq.~(\ref{a13})
\begin{equation}
\lim_{\qq\to\infty}\bar E_{max} (q ,\bar y_0,p)=\bar y_0+p \ ,
\label{a63}
\end{equation}
i.e., the asymptotic limit of the maximum excitation energy is finite. 
Apart from the order of the $p,E$ integration, Eq. (\ref{a60})  is
identical to the result of Ref. \cite{cio}. 

There actually is no difficulty in computing the residue of the 
above neglected negative energy pole in the IA. However,  the same for 
higher order FSI terms seriously complicates a 3-dimensional reduction. 
Rather than  elaborating this point, we proceed towards our major goal,
namely the 3-dimensional reduction of the relativistic GRS series.

\subsection{Reduction of the relativistic GRS series.}

We thus consider the relativistic GRS series Eqs.~(\ref{a49}), (\ref{a50})
and start with its lowest order term 
$W_0^{GRS}=-(1/\pi){\rm Im}[\Gamma_A G_N \tilde G_{1,A} G_N\Gamma_A]$. 
Applying the  above non-relativistic limit one obtains
\begin{eqnarray}
W_0^{GRS}(\nu,\bmq )&=&-\frac{1}{\pi}{\rm Im} \sum_n
\int{d^3p\over (2\pi)^3} \frac{dp_0}{2\pi}
\frac {i(8M_{A-1}M^2)^{-1}
|\langle\Gamma_A(\bmp )|\Phi_{A-1}^{(n)}\rangle |^2}{
\left (M_A-p_0-M-{\bmp^2\over 2M}+i\eta\right )^2(p_0-M_{A-1}-{\cal E}_n+i\eta )}
\nonumber\\*[5pt]
&&{1\over 2(M_A-p_0)\nu -2\bmp\bmq -Q^2+i\eta } \ .
\label{a64}
\end{eqnarray}
In contrast to the relativistic IS, the modified propagator $\tilde G_N$
has only one pole in the lower half of the complex $p_0$ plane and simple
calculus produces
\begin{eqnarray}
W_0^{GRS}(\nu,\bmq )=\sum_n\int  {d^3p\over (2\pi )^3} |\varphi_A^{(n)}(\bmp )|^2
\delta \left  [2\left (M-\Delta -{\cal E}_n\right ) \nu -
2\bmp\bmq -Q^2 \right ] \ .
\label{a65}
\end{eqnarray}
Integration over $\cos (\bmp ,\bmq )$ then yields
\begin{equation}
{\cal F}^{GRS}_{0}(y_G,\bmq )={1\over 4\pi^2} \bigg [
\int\limits_{|y_G|}^{\infty}pdp
\int\limits_0^{\tilde E_{max}}
{\cal P}(p,E)dE+\theta (y_G)\int\limits_0^{y_G}pdp\int
\limits_{\tilde E_{min}}^{\tilde E_{max}}
{\cal P}(p,E)dE\bigg ]\ ,
\label{a66}
\end{equation}
where
\begin{equation}
y_G=\frac{M}{\qq}\bigg \lbrack
\nu\bigg (1-\frac{\Delta}{M} \bigg )-\frac{Q^2}{2M}\bigg \rbrack
\label{a67}
\end{equation}
is a relativistic generalization of $y_W$,  Eq.~(\ref{a24}), derived in 
Refs. \cite{sag1,sag2}. The integration limits in (\ref{a66}) are 
\begin{equation}
\tilde E_{max\atop min}(q,y_G,p)={(y_G\pm p)\qq\over\nu} \ .
\label{a68}
\end{equation}
In particular
\begin{equation}
\lim_{\qq\to\infty}\tilde E_{max} (q , y_G,p)=y_G+p
\label{a69}
\end{equation}
Since $y_G\to \bar y_0$ in the asymptotic limit, the above 
$\tilde E_{max}$ 
and its analog $\bar E_{max}$ in the IA, Eq.~(\ref{a63}), coincide. 

Eq. (\ref{a66}), the first term of the relativistic GRS series, is seen to  
contain the spectral function Eq.~(\ref{a8}) It 
does not resemble its non-relativistic counter-part, Eq.~(\ref{a26}), which 
contains exclusively the momentum distribution  $n(p)$. 
The latter is due to
the independence of the non-relativistic  nucleon propagator
$\tilde G_N$, Eq.~(\ref{a23}) on $p_0$, in contrast to
its relativistic counterpart, Eqs.~(\ref{a50}), (\ref{a51}).
The upper limit $\tilde E_{max}$, Eq.~(\ref{a68}), is  always 
finite. Consequently  Eq. (\ref{a26}) is not the non-relativistic 
limit of ${\cal F}_0^{GRS}$, Eq.~(\ref{a66}).
In  fact, it resembles more the
corresponding expression for the IA, Eq.~(\ref{a60}). 

Since the momentum distribution, $n(p)$,  Eq.~(\ref{a27}) is a
simpler function than the spectral function ${\cal P}(p,E)$
which depends on two variables, it is of practical interest to compare
expressions for the maximum excitation energy of the spectator.
One thus concludes from Eqs.~(\ref{a62}) and (\ref{a68}) that in
the non-relativistic regime 
$\nu\ll M$, $E^{GRS}_{max} =\tilde E_{max}\gg E^{IA}_{max}$. 
The replacement $\tilde E_{max} \to\infty$, 
and consequently ${\cal P}(p,E)$ by $n(p)$ in Eq.~(\ref{a66})), is 
therefore less of an offense in the GRS case, than 
$\bar E_{max}=E^{IA}_{max}$ $\to \infty$  
in the IA  expression  Eq.~(\ref{a60}).

We now turn to  FSI terms  in  the relativistic 
GRS series, Eq.~(\ref{a49}), for instance the dominant FSI term
\begin{eqnarray}
W_1^{GRS}(\nu,\bmq )&=&-\frac{1}{\pi}{\rm Im} \sum_n
\int {d^4pd^4p'\over (2\pi)^8} 
{(8M^2M_{A-1})^{-1}\langle\Gamma_A(\bmp )|\Phi_{A-1}^{(n)}\rangle \over
\left (M_A-p_0-M-{\bmp^2\over 2M}+i\eta\right )(p_0-M_{A-1}-{\cal E}_n+i\eta
)}
\nonumber\\*[5pt]
&&{[2\nu (p'_0-p_0)-2\qq (p'_z-p_z)]V_{1;nn'}(\bmp -\bmp' )\over
[2(M_A-p_0)\nu -2p_z\qq -Q^2+i\eta ][2(M_A-p'_0)\nu -2p'_z\qq -Q^2+i\eta
]^2}
\nonumber\\*[5pt]
&&{\langle\Phi_{A-1}^{(n')}|\Gamma_A(\bmp' )\rangle \over
\left (p'_0-M_{A-1}-{\cal E}_{n'}+i\eta\right )\left (M_A-p'_0-M
-{{\bmp'}^2\over 2M}+i\eta\right )} \ .
\label{a70}
\end{eqnarray}
As in Eq. (\ref{a64}) for $W_0^{GRS}$, one reduces $W_1^{GRS}$, 
Eq. (\ref{a70}), by performing the $p_0,p'_0$ integrations 
over the isolated spectator poles and the result for the corresponding
reduced structure function Eq. (\ref{a59}) becomes
\begin{eqnarray}
{\cal F}^{GRS}_1(y_G, \bmq )=-{1\over \pi}
{\mbox{Im}} \sum_{nn'}\int\frac {d^3\bmp d^3\bmp'}{(2\pi)^6}
{\varphi_A^{(n)}(p)\left [{\displaystyle\nu\over \qq}({\cal E}_n-{\cal E}_{n'})
-(p_z-p'_z)\right ]V_{1;nn'}(\bmp -\bmp' )\varphi_A^{(n')}(p')
\over \left (y_G-p_z-{\displaystyle\nu\over \qq}{\cal E}_n+i\eta \right )
\left (y_G-p'_z-{\displaystyle\nu\over \qq}{\cal E}_{n'}+i\eta
\right)^2}\ .
\label{a71}
\end{eqnarray}
Upon neglect of the small  relativistic corrections 
$(\nu/\qq)({\cal E}_n-{\cal E}_{n'})$ in the numerator, one compares
Eq.~(\ref{a71}) with its non-relativistic analog Eq.~(\ref{a30a}).
The latter turns into the former
upon the following replacements of the scaling variable
and Green's function of the recoiling nucleon, Eq.~(\ref{a23})
\begin{eqnarray}
&&y_W \to y_G
\nonumber\\
&&\tilde G_N \to \tilde G^r_N(\nu,\bmq,\bmp )=
\left ({M\over \qq}\right ) 
{1\over y_G-p_z-{\displaystyle\nu\over \qq}{\cal E}_n+i\eta}\ .
\label{a72}
\end{eqnarray}

At this point we return to the non-relativistic kinetic energy
$p^2/2M_{A-1}$ which   has been neglected above
and is valid for all but the lightest spectators.
It is actually straightforward to include that energy,  which amounts to
replacing  
$y_G$, Eq. (\ref{a67}), by the $A$-dependent scaling variable
\begin{equation}
y_G^A=\frac {2y_G}{1+\sqrt{1+(2\nu y_G/ M_{A-1}\qq)}} \ .
\label{a73}
\end{equation}

One notes that the energy shift   
$(\nu /\qq){\cal E}_n$ in the
propagator, $\tilde G_N$, Eq.~(\ref{a72}), 
puts a finite upper limit to the maximum excitation
energy of the spectator in  ${\cal F}_1^{GRS}$, Eq.~(\ref{a71}).  
The same has been discussed for  the lowest order
term ${\cal F}_0^{GRS}$, Eq.~(\ref{a66}), and occurs in all higher
FSI terms. Those energy shifts should therefore be retained in the denominator
of Eq.~(\ref{a71}).   We neglected however their differences 
in the numerator of the same equation.

The above 3-dimensional reduction can be extended straightforwardly to all 
higher order terms of the relativistic GRS series leading to the result 
\begin{equation}
{\cal F}^{GRS}(y_G^A,\qq)=\sum_{j=0}^\infty \left ({M\over\bmq }\right )^j
{\cal F}^{GRS}_j(y_G^A,\qq ) \ ,
\label{a74}
\end{equation}
It obviously differs from its non-relativistic analog  Eq.~(\ref{a25b})
by the $q$-dependence of its expansion coefficients,  which is due 
to the ${\displaystyle\nu\over \qq}{\cal E}_n$ term in Eq.~(\ref{a72}).

A special case is the deuteron target $D$ for which   ${\cal E}_n$=0. 
This enables to  reinstate in Eq.~(\ref{a74}) the  $q$-independent expansion 
coefficients  ${\cal F}_j^{GRS}(y_G^D,\qq )\equiv {\cal F}_j^{GRS}(y_G^D)$
and the construction of the
reduced relativistic structure  function Eq.~(\ref{a74})   from the  
non-relativistic one  (cf. Eqs.~(\ref{a9}), ({\ref{a25}))
\begin{eqnarray}
{\cal F}^{REL}(y_G^D,\qq)=F^{NR}(y_W\to y_G^D,\qq )\ ,
\label{a75}
\end{eqnarray}
or alternatively
\begin{mathletters}
\label{a76}
\begin{eqnarray}
&&W^{REL}(\nu,\bmq )={\nu\over \tilde\nu}W^{NR}(\tilde\nu,\bmq )
\label{a76a}\\*[5pt]
&&\tilde\nu=\left (1-{\Delta\over M}+{\nu\over 2M}-{(y_G^D)^2\over
2M^2}
\right )\nu \ .
\label{a76b}
\end{eqnarray}
\end{mathletters}
Eq.~(\ref{a75}) implies that a calculation of the FSI part of ${\cal F}$
requires the 
solution of a 3-dimensional, instead of a more complicated 4-dimensional 
scattering equation.

It would be desirable to reach a similar simplification for
targets with $A\ge 3$. A hint how to proceed comes from a
comparison of the dominant FSI term $F_1$, Eq.~(\ref{a30a}), of 
the non-relativistic GRS series with the summed FSI, Eq.~(\ref{a33}).

It has been shown in Appendix A that for the non-relativistic case, the
summed FSI expression  (the second term in Eq.~(\ref{a33})) is obtained from 
$F_1$, Eq.~(\ref{a30a}) by $V_1\to T$  and addition of 
$M\Delta_n(\bmp)/\qq$ to the propagators.                                        
On account of the similarity of $F_1$ and 
${\cal F}_1$ (cf. Eqs.~(\ref{a30a}) and (\ref{a71})), we conjecture that the 
relativistic summed FSI  are similarly generated
from ${\cal F}_1^{GRS}$, Eq. (\ref{a71}). The final result is
\begin{eqnarray}
&&{\cal F}^{GRS}(y_G,\bmq )={\cal F}_0^{GRS}(y_G,\bmq )-\frac{M}{ \pi\qq }
{\mbox{Im}} \sum_{n,n'}\int\frac {d^3\bmp d^3\bmp'}{(2\pi)^6}
\nonumber\\*[5pt]
&&\frac {\varphi_A^{(n)}(\bmp )(p'_z-p_z)T_{nn'}
(\tilde E_{N,A-1};\bmp +\bmq ,\bmp' +\bmq )
\varphi_A^{(n')}(\bmp' )}    
{\left (y_G-p_z-{{\displaystyle\nu} {\cal E}_n\over \qq}
-{M\Delta_n(\bmp )\over\qq} +i\eta\right )
\left (y_G-p'_z-{{\displaystyle\nu} {\cal E}_{n'}\over \qq}
-{M\Delta_{n'}(\bmp' )\over\qq}+i\eta\right )
(y_G-p'_z-{{\displaystyle\nu} {\cal E}_{n'}\over \qq}+i\eta)}\ .
\label{a77}
\end{eqnarray}
We emphasize  again the occurrence of 3-dimensional 
$N$-spectator transition amplitudes $T_{nn'}$ for off-shell energy 
(cf. Eq. (\ref{a35}))
\begin{equation}
\tilde E_{N,A-1}={(y_G+\qq )^2\over 2M}-\Delta -{p^2\over 2M} \ .
\label{a78}
\end{equation}

The occurrence of a $T$ operator usually indicates summation of GRS terms 
in $V_1$, which is mandatory if the latter is singular.
As was the case for the 
non-relativistic case Eq.~(\ref{a33}),
the relativistic expression for the structure function 
in terms of the
scattering operator $T$, Eq.~(\ref{a77}) is more general that
the entire GRS series. 

Eqs.~(\ref{a75}) and  (\ref{a77}) are our main results. 
They are the outcome of  very accurate 3-dimensional 
reduction of the
relativistic structure functions of targets composed of
point-particles. The reduction is a direct consequence of the
separation in slow and fast target nucleons.
Compared with a non-relativistic GRS theory, FSI 
interactions are summed by means of a 3-dimensional scattering operator. 
Relativistic effects are only manifest in a relativistic scaling 
variable and in an additional energy shift in $N$ propagators.

In spite of the role  $(\nu/\qq){\cal E}_n$ plays in the
the limits of the excitation energies of the spectator, we wish to explore 
closure over those excitations in Eq. (\ref{a77}), replacing state-dependent
quantities by suitable averages, ${\cal E}_n\to \langle {\cal
E}\rangle$. This  leads to   an operator
$T$, which  describes the scattering of nucleon $'1'$ from $A-1$, 
fixed 
spectator nucleons (see for instance Ref. \cite{rj}). In particular one may
expand $T$ in a Watson series of scattering operators $t$ for nucleon
pairs and retain only the lowest order term. The result is 
\begin{eqnarray}
&&{\cal F}^{GRS}(y_G,\bmq )\approx {\cal F}_0^{GRS}(y_G,\bmq )-
\frac{M}{ \pi\qq }{\mbox{Im}} 
\int\frac {d^3\bmp d^3\bmp'}{(2\pi)^6}\rho_2(\bmp,\bmp';\bmq)
\nonumber\\*[5pt]
&&\frac {(p'_z-p_z)
\langle \bmp,-\bmp+\bmq|t(\tilde E_{N,A-1}|\bmp',-\bmp' +\bmq )\rangle }
{\left (y_G-p_z-{{\displaystyle\nu} \langle {\cal E}\rangle \over \qq}
-{M\langle \Delta(\bmp )\rangle \over\qq} +i\eta\right )
\left (y_G-p'_z-{{\displaystyle\nu} \langle {\cal E}\rangle \over \qq}
-{M\langle \Delta(\bmp' )\rangle \over\qq}+i\eta\right )
(y_G-p'_z-{{\displaystyle\nu} \langle {\cal E}\rangle \over
  \qq}+i\eta)}\ .
\label{a79}
\end{eqnarray}
Above $\rho_2(\bmp,\bmp',\bmq)$ is the two-particle density 
matrix in the
momentum representation, i.e. the Fourier transform of the same 
in coordinate representation (cf. Eq. (\ref{a30b})).

Clearly, for sufficiently high energy transfers $\nu$, nucleons may be excited, 
and this is manifest in  the nucleon structure functions  $F^N$.
Those dynamical features
should be built into a theory  and an example is a generalized
convolution of structure functions for nucleons and for a
nucleus, composed of point-nucleons\cite{gr}.
\begin{eqnarray}
F_2^A(x,Q^2)=
\int_x^A dz\ f^{PN}(z,Q^2) F_2^N \bigg (\frac{x}{z},Q^2 \bigg ) \ .
\label{a80}
\end{eqnarray}
Here  $f^{PN}(x,Q^2)$ is  ${\cal F}^{GRS}(q,y_G^A)$, 
Eqs.~(\ref{a73}), (\ref{a77}), expressed in the Bjorken variable $x$ and
$Q^2$. 
The above Eqs.~(\ref{a79}), (\ref{a80}) comes close to the expression used 
in actual calculations of nuclear structure functions\cite{rt}.

We conclude this Section by stating that for the same reasons 
as for the non-relativistic case forwarded at the end of Section II,
the relativistic GRS is expected to show better convergence than the IS
series. We shall provide  proof, using inclusive scattering  of  3.595
GeV electrons on $^4$He, for which the nuclear input can be computed
with high precision. Fig.~2 thus shows 
cross sections for two scattering angles $\theta=25^{\circ}, 30^{\circ}$  
as function of the energy loss, in a standard way related to the
$^4$He structure function, Eq.~(\ref{a80}).
\vskip1cm
%%%%%%%%%%%%%%%%%%%%%%%%%%%%%%%%%%%%%%%%%%%%%%%%%%%%%%%%%%%%%%%%%%%%%%%%%
\begin{minipage}{13cm}
\begin{center}
\leavevmode
\epsfxsize=10cm
\epsffile{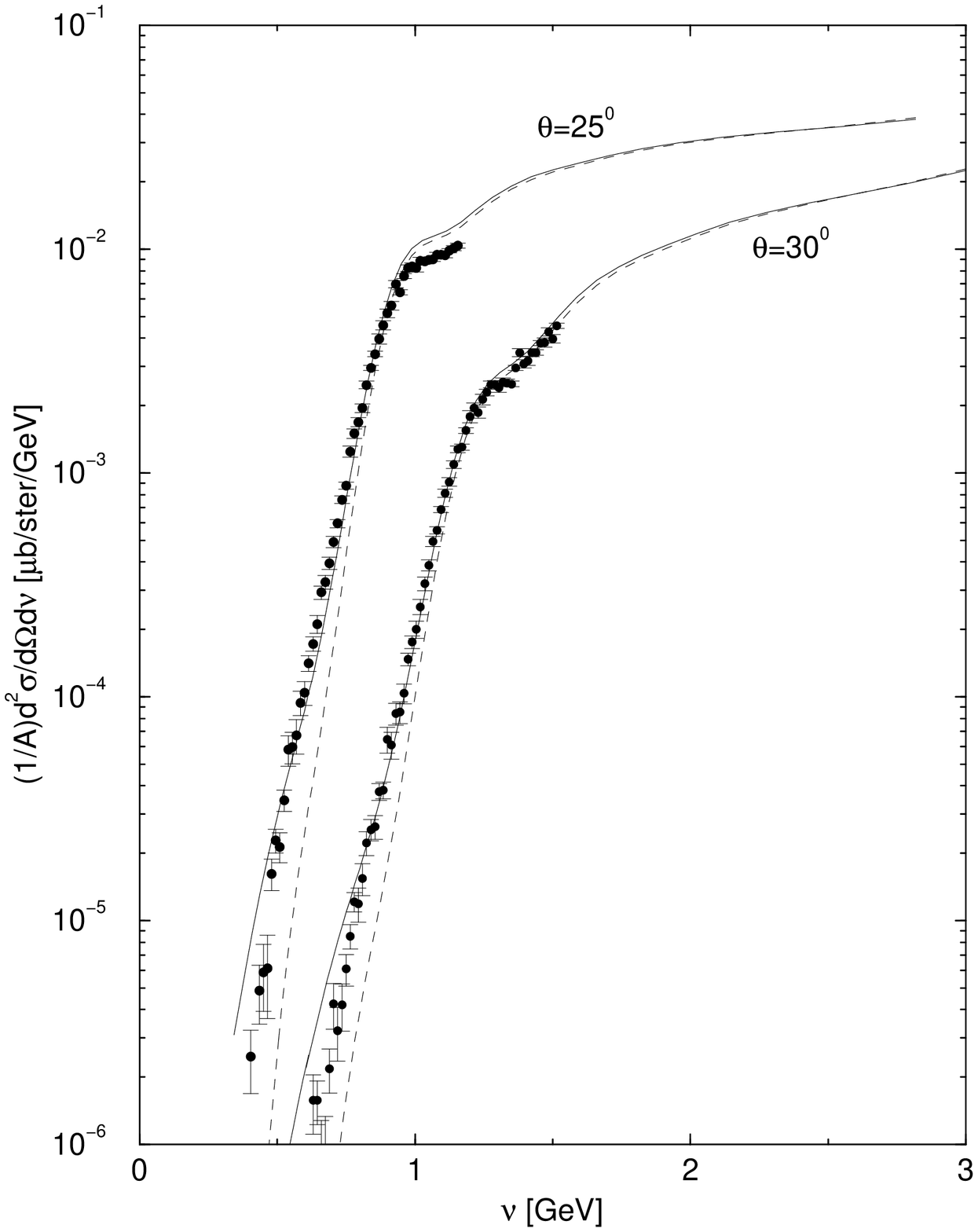}
\end{center}
{\begin{small}
Fig.~2. Double-differential cross-section for $e+^4He\to e+X$   
inclusive scattering as a function of $\nu$. The dashed line      
is the IA and the solid line corresponds to the lowest term
of the relativistic GRS series.                                   
\end{small}}
\end{minipage} \\ \\
%%%%%%%%%%%%%%%%%%%%%%%%%%%%%%%%%%%%%%%%%%%%%%%%%%%%%%%%%%%%%%%%%%%%%%%%%

Inspection shows that, except for the smallest $\nu$, the drawn lines
representing the lowest order GRS
prediction using  Eq.~(\ref{a66}) nearly accounts for the data\cite{day}.  
In contrast the dashes for the IA based on Eq.~(\ref{a60}) show sizeable       
discrepancies. Details can be found in Ref.\cite{vkr}. Similar evidence 
comes from $D$ data\cite{sag1,rtd}.

\section{Summary}

In this paper we studied a relativistic GRS series for structure functions
of a nucleus composed of point-nucleons. The latter we simplified,
exploiting  non-relativistic
features of all quantities there, which are related  to slow target
nucleons,
and only treated relativistically  the nucleon which absorbs the high
momentum of the transferred photon in inclusive scattering. Our focus is on
an accurate 3-dimensional reduction of the expression, which is possible
through a specific feature of a modified nucleon propagator, namely its
linear
momentum dependence. This is in contrast with the standard relativistic IS.

In the case of a  deuteron target, the above 3-dimensional reduction
leads to a perfect correspondence between the relativistic and the
corresponding non-relativistic expressions for   its structure 
function.
 The derivation of that  mapping does not employ
light-cone kinematics which has similar features, but
without the need to relax spherical symmetry  \cite{karm}.

For targets with $A\ge 3$ no such mapping can be proved rigorously. We
emphasized though the close correspondence between the non-relativistic and
relativistic dominant FSI terms  and  then conjectured, that the
same correspondence holds between   the non-relativistic and relativistic
summed FSI contributions. The latter  can  then
be calculated, using 3- instead of a  more complicated
4-dimensional scattering equation. The above rests on the assumption that
the driving term of  the 4-dimensional Bethe-Salpeter scattering equation is
local and given by a sum of pair interactions, as has also been assumed in
the non-relativistic case. 

Our final remarks regarded the application of the obtained results. In spite of
the proved reduction and correspondence, it is still non-trivial to
solve multi-channel scattering of a nucleon from a fully interacting $A$-1
nucleon spectator. We mentioned the  approximation, where
many-body  transition operators are replaced by a sum of scattering
operators for a pair of nucleons. In addition we recalled
the incorporation of nucleons with internal dynamics.
Both features are about  the basis of previously performed computations \cite{rt}.

\section{Acknowledgments.}
ASR is grateful to Byron Jennings for having cooperated in an, initially
different approach and having later commented on the present one. He also 
much profited from the constructive criticism of Roland Rosenfelder.
SAG thanks Cyclotron Institute at Texas A$\&$M University for kind
hospitality.

\appendix
\section{Final State Interaction in GRS expansion.}

In the following we expand on the derivation of Eq.~(\ref{a31})
for the summed FSI contribution, which was previously given 
in Ref.\cite{sag1}.  Consider non-relativistic structure function given by
the GRS series, Eqs.~(\ref{a25a}). Using Eq.~(\ref{a20}) (with $\tilde
G_{1,A}
\equiv \tilde G_N$, Eq.~(\ref{a23})), we can rewrite the FSI part of the
structure function as 
\begin{equation}
W_{FSI}(\nu ,\bmq )=-\frac{1}{\pi}{\rm Im}
\langle \Phi_A|\sum_{n=1}^\infty\tilde G_N(\tilde V\tilde G_N)^n|\Phi_A\rangle
= -\frac{1}{\pi}{\rm Im}
\langle \Phi_A|\tilde G_N\tilde VG_{1,A}|\Phi_A\rangle \ , 
\label{ap1}
\end{equation}
where $G_{1,A}\equiv G_{1,A}(E_A+\nu ,\bmq )$, is the total Green's
function after the absorption of the virtual photon, Eq.~(\ref{a2}).
Using the Lippmann-Schwinger equation (\ref{a16}) one can express 
$G_{1,A}$ in terms of the scattering operator $T$
\begin{equation}
G_{1,A}=(1+G_{1,A}^{(0)}T)G_{1,A}^{(0)}
\label{ap2}
\end{equation}
where $G_{1,A}^{(0)}\equiv G_{1,A}^{(0)}(E_A+\nu ,\bmq )$, Eq.~(\ref{a3}).
Substituting Eq.~(\ref{ap2}) into Eq.~(\ref{ap1}) and using
$\tilde V=H_A-E_A=V_1-g_0^{-1}$ with $g_0\equiv G_{1,A}^{(0)}(E_A,0)$, 
we obtain  
\begin{equation}
W_{FSI}=-\frac{1}{\pi}{\rm Im}
\langle \Phi_A|\tilde G_N(V_1-g_0^{-1})
(1+G_{1,A}^{(0)}T)G_{1,A}^{(0)}|\Phi_A\rangle \ ,
\label{ap3}
\end{equation}
One easily finds from Eqs.~(\ref{a21}), (\ref{a23}) that 
$\tilde G_N-G_{1,A}^{(0)}=\tilde G_Ng_0^{-1}G_{1,A}^{(0)}$. Then using 
$V_1(1+G_{1,A}^{(0)}T)=T$ we can rewrite Eq.~(\ref{ap3}) as 
\begin{equation}
W_{FSI}=-\frac{1}{\pi}{\rm Im}
\langle \Phi_A|G_{1,A}^{(0)}TG_{1,A}^{(0)}-g_0^{-1}
\tilde G_NG_{1,A}^{(0)}|\Phi_A\rangle
\ .  
\label{ap4}
\end{equation}

At the last step we use the following relation between the scattering 
operator $T$ and the target wave function, valid for any local interactions
\begin{equation}
\langle \Phi_A|g_0^{-1}=\langle \Phi_A|G_{1,A}^{(0)}\tilde G_N^{-1}T
\label{ap5}
\end{equation}
The relation can easily obtained by multiplying the Lippmann-Schwinger equation
$T=V_1(1+G_{1,A}^{(0)}T)$ by $\langle \Phi_A|$ and using the Schr\"odinger
equation for the target wave function: $\langle \Phi_A|V_1= \langle
\Phi_A|g_0^{-1}$. Then Eq.~(\ref{ap4}) can be rewritten as
\begin{eqnarray}
W_{FSI}&=&-\frac{1}{\pi}{\rm Im}
\langle \Phi_A|G_{1,A}^{(0)}TG_{1,A}^{(0)}
-G_{1,A}^{(0)}\tilde G_N^{-1}T\tilde G_NG_{1,A}^{(0)}
|\Phi_A\rangle\nonumber\\*[5pt]
&=&-\frac{1}{\pi}{\rm Im}
\langle \Phi_A|G_{1,A}^{(0)}\tilde G_N^{-1}[\tilde G_N,T]
G_{1,A}^{(0)}|\Phi_A\rangle 
\ ,
\label{ap6}
\end{eqnarray}
thus obtaining the desired expression for the summed FSI contribution in
the GRS series.


\begin{references}
 
\bibitem{cio}
See for instance: C. Ciofi degli Atti, E.Pace and G. Salme, Phys. Rev.
 C43, 11275 (1991); C. Ciofi degli Atti, D.B. Day and S. Liutti,
$ibid$ C46, 1045 (1994).
 
\bibitem {om1}
O. Benhar, A. Fabrocini, S. Fantoni, G.A. Miller, V.R. Pandharipande and
I. Sick,  Phys. Rev.  C44 , 2328 (1991); Phys. Lett. B359, 8 (1995).
 
\bibitem{os}
P. Fernandez de Cordoba, E. Marco, H. Mutter, E. Oset and A. Faessler,
Nucl. Phys. A 611, 514 (1996).
 
\bibitem{cio1}
C. Ciofi degli Atti and S. Simula, Phys. Lett. B325, 276 (1994).
 
\bibitem{koh}
A. Kohama, K. Yazaki and R. Seki, Nucl. Phys. A 662,175 (2000).

\bibitem{braun}
M. Braun, C. Ciofi degli Atti and D. Treleani, Phys. Rev. C 62, 034606
(2000).
 
\bibitem{grs}
H.A. Gersch, L.J. Rodriguez and Phil N. Smith, Phys. Rev.
A5, 1547 (1973).

\bibitem{retal}
A.S. Rinat, M.F. Taragin, F. Mazzanti and A. Polls, Phys. Rev. B 57, 5347
(1998), and references therein.
 
\bibitem{west}
G.B.  West, Phys. Rep.  18, 264 (1975).
 
\bibitem{pg}
N. Poliatzki and S.A. Gurvitz, Nucl. Phys. A524, 217 (1991).
                                                                         
\bibitem{gr1}
S.A. Gurvitz and A.S. Rinat, Phys. Rev. C47, 2901 (1993).
 
\bibitem{sag1}
S.A. Gurvitz, Phys. Rev. C42, 2653 (1990).
 
\bibitem{sag2}
S.A. Gurvitz, Phys. Rev. D52, 1433 (1995).

\bibitem{mul} 
P.J. Mulders, Phys. Rep. 185, 83 (1990).

\bibitem{rj}
A.S. Rinat and B.K. Jennings, Phys. Rev. C59, 3371 (1999).

\bibitem{gr}
S.A. Gurvitz and A.S. Rinat, Progress in
Nuclear and Particle Physics,  34, 245 (1995).

\bibitem{rt}
A.S. Rinat and M.F. Taragin, Nucl. Phys. A598, 349 (1996); $ibid$
A620, 412 (1997); Erratum: $ibid$ A623, 773 (1997); Phys. Rev.
C60, 044601 (1999).

\bibitem{day} 
D.B. Day {em et al}, Phys. Rev. C48, 1849 (1993). 

\bibitem{vkr} 
M. Viviani, A. Kievsky and A.S. Rinat, To be submitted to 
Phys. Rev. C.

\bibitem{rtd}
A.S. Rinat and M.F. Taragin, in preparation.

\bibitem{karm} J. Carbonell, B. Desplanques, V.A. Karmanov and J.F.
Mathiot, Phys. Rep.  300, 215 (1998).
 
\end{references}
\end{document}